\documentclass[12pt,a4paper]{article}
\usepackage[margin=1.0in]{geometry}
\usepackage{times}
\usepackage{setspace}
\usepackage{graphicx}
\usepackage{amsmath,amssymb,amsthm}
\usepackage{hyperref}
\usepackage{float}
\usepackage{caption}
\usepackage{lipsum}

\newcommand{\remove}[1]{}

\usepackage{color}

\makeatletter
\partopsep\z@ \textfloatsep 10pt plus 1pt minus 4pt
\def\section{\@startsection {section}{1}{\z@}{-3.0ex plus -0.5ex minus
-.2ex}{2.3ex plus .2ex}{\large\bf}}
\def\subsection{\@startsection{subsection}{2}{\z@}{-2.5ex plus -0.5ex
minus -.2ex}{1.5ex plus .2ex}{\normalsize\bf}}
\def\@fnsymbol#1{\ensuremath{\ifcase#1\or 1\or 2\or
    3\or 4\or 5\or 6\or 7 \or 8\ or 9 \or 10\or 11 \else\@ctrerr\fi}}
\makeatother

\let\OLDthebibliography\thebibliography
\renewcommand\thebibliography[1]{
  \OLDthebibliography{#1}
  \setlength{\parskip}{0pt}
  \setlength{\itemsep}{0pt plus 0.3ex}
}

\begin{document}

\begin{center}
 {\huge\bf A National Effort for Motivating Indian}
\vskip 0.05in
{\huge\bf  Students and Teachers }
\vskip 0.07in
{\huge\bf  towards Algorithmic Research}
      ~\\
      ~\\
      ~\\

Subir Kumar Ghosh\\
School of Technology and Computer Science\\ 
Tata Institute of Fundamental Research\\ 
Mumbai 400005, India\\ 
ghosh@tifr.res.in 

\vskip 0.2in 
Sudebkumar Prasant Pal\\
Department of Computer Science and Engineering\\
Indian Institute of Technology\\ Kharagpur 721302, India\\
spp@cse.iitkgp.ernet.in 
\end{center} 

\begin{abstract}

During 2008-2015, twenty-two introductory workshops 
  on graph and geometric algorithms  were
organized for teachers and students (undergraduate, post-graduate and doctoral) 
of engineering colleges and universities at 
different states and union territories of India.
The lectures were meant to provide
exposure to the field 
of graph and geometric algorithms and to motivate the participants towards
research.
Fifty-eight professors
from TIFR, IITs, IISc, IMSc, CMI, ISI Kolkata,  and other institutes and
universities delivered invited lectures on different
topics in the design and analysis of algorithms, discrete applied mathematics,
computer graphics, computer vision,  and robotics.    
The first four workshops were funded by TIFR, BRNS and IIT Kharagpur, and 
the remaining
workshops were funded by the NBHM.
In this paper, we present the salient features of these 
workshops, and state our observations on the national impact of these workshops.

\end{abstract}

\section{Introduction}

\subsection{Background}


Ab\={u} Jafar Muhammad al-Khw\={a}rizm\={\i}, a Persian astronomer and
mathematician, wrote a treatise in 825 AD, 
{\it Kit\={a}b his\={a}b al-adad al-\d{h}ind\={\i}} ({\it Book on Calculation 
with Hindu Numerals}), which was translated into Latin in the early 12th 
century as {\it Liber Algorismi de numero Indorum} ({\it The Book of Algorismi 
on Indian Numerals}).
The word ``Algorism''-- the Latin form of al-Khw\={a}rizm\={\i}'s name -- came to be
applied to any systematic work on ancient Indian-style computational mathematics.
The present term ``algorithm'' is a distorted form of ``algorism''
\cite{ad2015}.

\medskip

The field of algorithms, which  is at the very heart of computer science,
has witnessed a number of significant advances during the last
five decades. 
These advances include the development of faster algorithms and the 
discovery of certain natural problems for which all known algorithms 
are inefficient \cite{gj1979}. These startling results have
kindled a keen interest in  the area of algorithm
design and analysis.
Teaching and research in this foundational aspect of computing is 
therefore a natural 
and desirable thrust area.

\medskip

As algorithms are also at the heart of every nontrivial computer 
application, 
computer scientists and professional programmers are expected to know about
the basic algorithmic toolbox: structures that allow efficient organization and retrieval
of data, frequently used algorithms and generic techniques for modeling, understanding
and solving algorithmic problems. Hence, algorithmic studies form a major component of 
computer science programs in colleges and universities.

\medskip

In the last four decades, graph and geometric problems have been studied
by computer science researchers using the framework of design and analysis
of algorithms \cite{ahu75,rt83}. 
While graph algorithms have been studied for 
almost 300 years, graphs provide essential models for many application areas of computer 
science, and at the same time, they are fascinating objects of study in pure and applied 
mathematics. There have been a number of exciting developments in graph theory that 
are important for designers of algorithms. Correspondingly, the algorithmic 
viewpoint of computer science has stimulated much research in graph theory. Graph theory and 
graph algorithms are inseparably intertwined subjects. 

\medskip 

On the other hand, the main impetus for the development of geometric  algorithms came 
from the progress in computer graphics, computer-aided design and 
manufacturing \cite{BKOS97}. In 
addition, algorithms are also designed for geometric problems that are classical in nature. 
The success of the field can  be explained from the beauty of the geometry problems studied,
the solutions  obtained, and by the many application domains -- computer graphics, geographic 
information systems, robotics and others, in which geometric algorithms play a crucial role.

\subsection{Motivation}

We know that basic algorithmic research has helped building 
today's computer technology
and will keep on playing a crucial role in the achievement of 
tomorrow's technological
breakthroughs. Therefore, we believe that algorithmic research is 
certainly important even for India.

\medskip

At present, research is being carried out in India in several sub-areas
of algorithms and results are being published in reputed conferences
and journals in computer science and discrete applied mathematics.
Algorithmic research started in India in premier academic institutes like
TIFR, IITs and
IISc, way back in the 70's. In recent years, algorithmic research is also
carried out in computer industries in India including Microsoft Research,
IBM Research, TCS Innovation Labs, etc. Today, computer industries
in India provide financial support to projects on algorithmic research
in Indian academic institutes and also provide fellowships to
Ph.D. students working in algorithmic research.

\medskip

The need for algorithmic research in India was felt in the
late 70's.  Perceiving future needs, the course ``Design and
Analysis of Algorithms" was introduced in the 80's in the undergraduate and
postgraduate levels as a core course
in premier institutes like IIT's, TIFR and IISc. This course used to cover
asymptotic analysis of running time and space requirements for 
quantitatively measuring the
efficiency of algorithms, searching and sorting algorithms, data structures,
sequential algorithms for graph and geometric problems,
NP-completeness, etc. 

\medskip

Designing efficient algorithms for graph and geometric problems
within the framework of the design and analysis of algorithms started 
also around the same time. There were a few researchers in India (including
S. N. Maheshwari (IIT Delhi), C. E. Veni Madhavan (IISc Bangalore), Subir
Ghosh (TIFR Mumbai)), who were involved in algorithmic research during that
time. In the 90's, Ph.D. and Master theses on algorithms for graph and 
geometric problems started coming up in increasing
numbers, demonstrating the acceptability of algorithmic research in India.
Many of these theses for designing efficient
algorithms used mathematical techniques from combinatorics,
probability theory, graph theory, etc.,  and algorithmic techniques 
like greedy methods, divide and conquer, approximations, randomization, etc.

\medskip

On the other hand, after personal computers became more and more available
in India in the late 80's, a large number of scientists, engineers and
administrators in India across all disciplines started using personal
computers for solving  problems arising in research, applications and 
in commerce.
This gave rise to the need of computer programmers, as the IT industries
proliferated. Naturally, several softwares were designed in 80's and 90's 
and many of  them used basic
algorithmic techniques and data structures that were routinely taught
in the courses of ``Design and Analysis of Algorithms" not only in 
IITs, IISc and TIFR but also in NITs, BITS Pilani, and several 
universities in India. 
Algorithmic studies
gained further impetus in the 90's in India like anywhere in the world
after personal computers became connected through networks, forming the 
world wide
web. New types of algorithms like streaming algorithms, web-based algorithms,
network algorithms, distributed algorithms, big data algorithms,
etc. came into play in the last two decades.

\medskip

Though 
the number of researchers working in India on graph and geometric algorithms
has increased substantially in the last two decades, the number of active researchers
in algorithms is still far lesser in India compared to the large number of
bright students and teachers
involved in studying and teaching computer science and discrete applied
mathematics. In
order to motivate them towards computer science research in general and
algorithmic research
in particular, twenty-two workshops of introductory lectures on graph and
geometric algorithms
were organized during 2008-2015 for teachers and students (undergraduate,
post-graduate, and
doctoral) of engineering colleges and universities at different locations in
India. 

\medskip
Our workshop series may be viewed as a human resource development program
for raising the
level of algorithmic knowledge amongst Indian college and university students
and teachers in
computer science and discrete applied mathematics.

\section{Workshop data}

\subsection{Organizational details}

\begin{figure}
\begin{center}
\includegraphics[width=1.0\columnwidth]{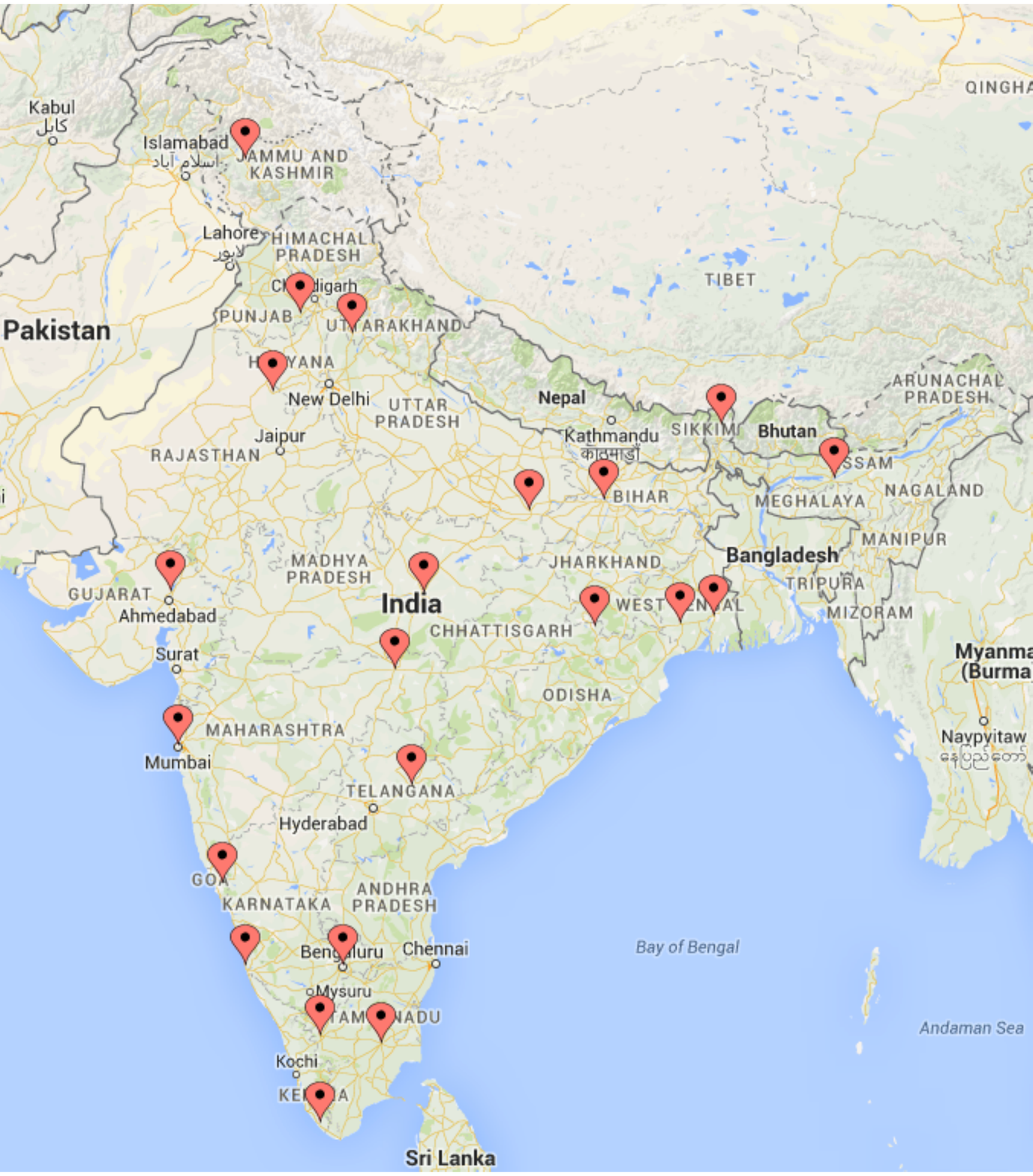}
\caption{Workshop venues} 
\label{workshopslocation} 
\end{center}
\end{figure}

During 2008-2015, twenty-two workshops of introductory lectures 
 on graph and geometric algorithms 
were organized   at different states and union territories
 of India as shown in Tables I(a) and II(b), and in  Figure \ref{workshopslocation}.
\newpage
\begin{center}
{\bf Table I(a): Dates, Workshop Venues and Coordinators}

\bigskip
\footnotesize
\begin{tabular}{| l| l |l| l|l|}
\hline
No&Date & Workshop Venue & Coordinators\\
\hline
1&July 22-23& Fr. Conceicao Rodrigues College of& Subir Ghosh, Sunil Surve, Deepak Bhoir\\ 
&2008&  Engineering, Mumbai, Maharashtra&  Vijay Bilolikar \\

\hline

2&October 31-  & Indian Institute of Technology& Subir Ghosh, Partha Bhowmick\\
&November 2 &Kharagpur, West Bengal& Sudebkumar Pal\\
&2008&&\\
\hline
3&January 22-24& Birla Institute of Technology \& Science&Subir Ghosh, Poonam Goel, Navneet Goel\\
&2009 & Pilani, Rajasthan&\\
\hline

4&July 15-18&  Indian Institute of Science & Subir Ghosh, Sathish Govindarajan\\  
 &2009  & Bangalore, Karnataka & Sunil Chandran, Vijay Natarajan\\
\hline

5&January 7-9  &  National Institute of Technology&  Subir Ghosh, Venkatesh Raman, Hemalatha\\
&2010& Tiruchirapalli, Tamilnadu  & Thiagarajan, A. Ramakalyan, Sunil Chandran\\ 

\hline
6&January 27-29& Banaras Hindu University& Subir Ghosh,  Arun Agrawal, Pramod\\  
&2010 &  Varanasi, Uttar Pradesh &  Mishra,  Sudebkumar  Pal, Sunil Chandran\\
\hline
7&March 25-27 & National Institute of Technology& Subir Ghosh, Banshidhar Majhi, Pankaj Sa\\
&2010 &Rourkela, Odisha &  Partha Goswami, Sudebkumar Pal\\  

\hline

8&28-30 October&   Thapar University &Subir Ghosh,  Deepak Garg, Sandeep Sen\\
&2010 & Patiala, Punjab&  Subhas Nandy\\ 
\hline

9&January 6-8& PSG College of Technology  &Subir Ghosh,   R. S. Lekshmi, R. Nadarajan\\  
&2011 &Coimbatore, Tamilnadu&     Venkatesh Raman\\

\hline

10&March 26-28&National Institute of Technology&  Subir Ghosh, Sawal Singh, Subhas Nandy\\
&2011&Patna, Bihar  &Sathish Govindarajan, Md. Haider\\
\hline

11&October 21-23& Indian Institute of Technology& Subir Ghosh,  R. Inkulu, Pinaki Mitra\\
&2011& Guwahati, Assam &   Sudebkumar Pal\\

\hline

12&January 10-12&National Institute of Technology & Subir Ghosh, Suresh Hegde, Shyam\\
&2012&Surathkal, Karnataka& Kamath,  P. Jidesh, Sathish Govindarajan\\ 

\hline

13&March 14-16& Dhirubhai Ambani Institute of  &Subir Ghosh,  Srikrishnan Divakaran, Rahul\\
&2012 & Information and Communication  &  Muthu,  Sathish Govindarajan,
V. Sunitha\\
&&Technology, Gandhinagar,  Gujarat&   \\
\hline

14&November 1-3& PDPM Indian Institute of Information &Subir Ghosh, Sudebkumar Pal, Pritee Khanna\\

&2012  &  Technology, Design and Manufacturing &Sraban Mohanty  \\
&&Jabalpur, Madhya Pradesh &\\

\hline

15&January 17-19& Birla Institute of Technology \& Science  &  Subir Ghosh,  Tarkeshwar Singh, Anil Pundir\\
&2013& Pilani, Goa&    B.M. Deshpande, Venkatesh Kamat\\

\hline

16&March  14-16 &Bengal Engineering and Science  & Subir Ghosh, Arindam Biswas, Sekhar Mandal\\
&2013& University, Shibpur, West Bengal &Partha Bhowmick, Chandan Giri, Sanjay Saha\\
\hline

17&October  23-25 & National Institute of Technology & Subir Ghosh, R.B.V. Subramanyam\\
&2013 & Warangal, Andhra Pradesh &  Sandip Das, P V Subba Reddy\\
\hline
18&January  23-25 & University of Kerala \& Indian Institute &Subir Ghosh, Manoj Changat, T.K. Manoj\\
&2014 &  of Information Technology  and  & Kumar, K. Satheesh Kumar, T. Radhakrishnan\\
&& Management, Trivandrum, Kerala &\\

\hline
19&March  6-8 & Indian Institute of Technology & Subir Ghosh, Sudebkumar  Pal\\
&2014& Roorkee, Uttarakhand & Rajdeep Niyogi, Vaskar Raychowdhury \\

\hline

\hline
20&October  16-18 &   Sikkim Manipal Institute of Technology  &Subir Ghosh, Sudebkumar  Pal,  Biswajit Deb\\ 
&2014 & \& Sikkim Government College, Sikkim &  Tejbanta Chingtham, Debabrata Purohit, \\

\hline
21&January 15-17 & Visvesvaraya National Institute of  & Subir Ghosh, Manish Kurhekar, Abhiram\\
&2015 & Technology, Nagpur, Maharashtra&  Ranade, Ravindra Keskar, Umesh Deshpande\\
\hline
22&May 18-20 & University of Kashmir & Subir Ghosh, Shariefuddin Pirzada\\
&2015 & Srinagar, Kashmir &Abhiram Ranade, S. M. K. Quadri\\
 
\hline

\end{tabular}
\end{center}


In addition to coordinators, 
two groups of people were involved in organizing these workshops.
One group typically consisted of students and staff of host institutes who were involved
essentially during the workshop days. The other group consisted of administrative 
and technical staff of
School of Technology and Computer Science, TIFR who were involved from the preparatory
phase till the completion of  every workshop spanning a period of over  3-4 months. 
Mr. Raymond J. D'Mello worked 
as the secretary for this
workshop series for the entire period of  2008-2015; he was assisted by 
Mr. John Barretto,  Mr. Pravin N. Bhuwad, Mr. W. K. Gawade and 
Mr. Nitin S. Gawandi.
Workshop websites for all the workshops during 2009-2015 were designed, developed, updated
and  maintained by Mr. Ravikumar Naik, 
and he also
provided a necessary platform for online registration of participants.

\subsection{Financial support}

The first four workshops were funded by Tata Institute of Fundamental Research 
(TIFR), Indian Institute of Technology, Kharagpur (IITKGP), and the Board of 
Research in Nuclear Science (BRNS)
of the Department of Atomic Energy, Government of India (see Table II(a)), and 
the remaining eighteen 
workshops were funded by the National Board for Higher Mathematics (NBHM), 
Department of Atomic
Energy, Government of India (see Table II(b)). There was no registration fee for the 
workshops after the first three workshops. Thus the workshops 
were almost entirely funded by 
the above mentioned organizations. 
In addition, some host institutes provided
small grants to cover some local expenses. In most cases, host institutes provided
many facilities including auditoriums and guest houses at subsidized rates. Some invited
speakers used their own travel grants for paying their  air fares. Taking
all these facts into consideration, it can be seen that  budgets for the workshops were 
indeed quite low.

\begin{center}
{\bf Table II(a): TIFR, BRNS and IITKGP  Support}

\bigskip
\footnotesize
\begin{tabular}{| l| l |l| l|}
\hline
Workshop Venue & TIFR Grant& BRNS Grant & IITKGP Grant\\

\hline
CRCE Mumbai& Rs. 80,000/-& --& --\\\hline
IIT Kharagpur& --&--&Rs. 2,15,000/-\\\hline
BITS Pilani& Rs. 3,00,000/-&--&--\\\hline
IISc Bangalore& Rs. 6,00,000/-&2,00,000/-&--\\     
\hline
\end{tabular}

\end{center}
\begin{center}
{\bf Table II(b): NBHM Support}

\bigskip
\footnotesize
\begin{tabular}{| l| l |l| l|}
\hline
Workshop Venue & NBHM Grant& Workshop Venue & NBHM Grant\\

\hline
BHU Varanasi& Rs. 2,60,000/-&IIITDM Jabalpur& Rs. 3,70,000/-\\
\hline
NIT Tiruchirapalli &Rs. 2,60,000/-&BITS Goa& Rs. 3,70,000/-\\
\hline
NIT Rourkela& Rs. 2,60,000/-& NIT Warangal& Rs. 3,70,000/-\\
\hline
Thapar University& Rs. 3,00,000/-&IIITM Kerala&Rs. 3,90,000/-\\
\hline
PSGTECH Coimbatore&Rs. 3,00,000/-&IIT Roorkee& Rs. 3,90,000/-\\
\hline
NIT Patna& Rs. 3,00,000/-& BESU Shibpur & Rs. 3,90,000/-\\
\hline
IIT Guwahati& Rs. 3,50,000/-& SMIT Sikkim &Rs. 4,10,000/-\\
\hline
NIT Karnataka&Rs. 3,50,000/-&VNIT Nagpur & Rs. 3,60,000/-\\
\hline
DAIICT GandhiNagar&RS. 3,50,000/-&University of Kashmir & Rs. 4,30,000/-\\

\hline
\end{tabular}

\end{center}

\subsection{ Invited lectures}

In the 22 workshops, 266 invited lectures were delivered by 
58 professors under 114 lecture titles.
A complete list is given in Appendix. 
These lectures can be classified broadly into four
categories: (i)  Algorithmic Paradigms, (ii) Graphs and Algorithms, (iii)
Geometry and Algorithms, and (iv) Geometry and Applications. 

\medskip

Under algorithmic
paradigms, introductory lectures were delivered on different types of 
algorithms \cite{csrl,kleinbergtardos}. Specifically, approximation 
algorithms \cite{VV03}, 
randomized algorithms \cite{motwani}, 
on-line algorithms \cite{br2005}, distributed algorithms \cite{Lyn96},
fixed parameter algorithms \cite{ni2006}, quantum algorithms \cite{NielsenChuang}
 and streaming algorithms \cite{mutu2005} were covered.

\medskip

Under graphs and algorithms, introductory lectures were delivered on graph theory 
\cite{B78,Mcg2004,fhh1969,JT95},
networks \cite{tripathy-ida:2013}, graph algorithms \cite{kleinbergtardos,lp2009} 
and scheduling \cite{l2004}. Specific topics covered from  graph theory were
graph matching, graph representations, extremal graphs,  graph 
partitioning, probabilistic graphs, topological minors of graphs, 
expander graphs, graph coloring, 
graph spanners,  and  
graph spectra. Similarly, specific topics covered from graph algorithms were 
planarity testing, network flow, shortest paths,
social networks and wireless networks. 
  
\medskip

Under geometry and algorithms, introductory lectures were delivered on computational geometry 
\cite{BKOS97,s-vaip-2007,PS1985},
combinatorial geometry \cite{E87,m2002} and digital geometry \cite{klette04}. 
Specific topics  of computational and combinatorial geometry
covered were geometric graphs, facility location, convexity of 
points, art gallery problems, duality transformation, geometric data 
structures, geometric shortest paths, geometric spanners, Voronoi diagrams,
centerpoint location, and geometric prune and search  technique.
Similarly, specific topics covered from digital geometry were shape transformations by 
local interchanges,
algorithms for circles and spheres in digital space, isothetic covers for 
digital objects, anatomies of lines
and circles in the light of number theory and on good digital distances.

\medskip
Under geometry and applications, introductory lectures were delivered on computer
graphics \cite{DeysBook,d2006,menache99}, computer vision \cite{mvg2004,Sali04}, visualization
\cite{HaJo2004} and robotics 
\cite{choset05,JCL:91}. Specific topics covered from these 
application areas were  3D reconstruction from images and videos, projective geometry
for graphics and computer vision, visibility maps for point clouds, 
motion structure using shape spaces, character animation, 
haptic rendering,  symmetry and structure detection for 3D
geometry, rendering using directional distance maps, cache
friendly compressed representation of geometry, 
modeling for shape classes,  manifold 
discovery in data, scalar field visualization,
3D surface reconstruction, graphics processor units, mesh generation,
object recognition,  a number
theoretic introduction to geometry,  and
sequential and online algorithms for robot path 
planning.

\subsection{Participants}

Participants of the workshops were mainly undergraduate, post-graduate and doctoral students and teachers of computer science and discrete applied mathematics. 
They were required to have taken basic courses in discrete mathematics, 
data structures and algorithms at the undergraduate level.
There were around fifty participants in the first workshop at CRCE Mumbai. Since then,
the number of participants were  at least 90  in every  workshop. In some workshops,
there were  around 120 participants. The workshop at IIT Guwahati had the maximum
number of 156 participants. 

\medskip

Normally, around three-fourths of the  participants
were from the host institute and its nearby institutes and university. 
The remaining participants, who were from the same state and from neighboring states,
were provided with return train fares and free boarding and lodging.
All participants were provided free lunch and snacks,
and were issued certificates for attending
workshops. Also, a standard textbook on algorithms (Indian edition) was gifted to every 
participant for further study.


\section{Impact on the participants}
\subsection{Interaction}

Usually, there were 4 or 5   invited one-hour lectures in a day. 
Many of these lectures were designed as 
interactive lectures. Around 10-20\% of the participants 
took active part in such interactions and many of them 
got inspired by this opportunity to involve themselves.
Interaction with speakers continued  
during lunch and tea breaks.
The active participants were mostly students from IITs, IISc, 
some NITs, BITS Pilani 
and ISI, Kolkata. 
Although around 15-30\% participants were really at a loss during many
of the lectures, the remaining participants followed several lectures 
even though 
they did not interact during the lecture.
In fact, many of them met the speakers during breaks and clarified doubts.
At the end of each workshop, we always had a feedback 
session. We received  appreciation and valuable comments in feedback sessions.
We also received emails from participants after almost every workshop on the 
merits and demerits of the workshop.

\medskip
In every workshop, there was a session on open problems in which 
unsolved research problems on algorithms were stated and discussed. 
Many participants showed interest in the sessions on open problems. 


\subsection{Panel discussion}

The general trend for computer science students in India is to take up a job in IT 
industry after completing B.Tech or M. Tech. degree. Most of them do not even
consider that research  in computer science could be his/her profession.  In view
of this, a panel discussion was conducted in every workshop on the 
topic ``Is research a good career option?'' Normally, some young invited speakers of 
the workshop
took the role of  panelists. During the discussions, participants asked
panelists  questions on various career options: 
What are the benefits in a research
profession? What are the pitfalls in a research career? How much salary one can expect in research profession?   
What are the good institutes in India for doing research? How can one get a good
research guide? Can good research be done
in an industry?  How do I convince my parents that research is a viable career option? 
Is it not better to do research in the USA rather than in India? 
During the panel discussions, extensive interactions took place due to diverse 
opinions and 
counter-opinions expressed by panelists and participants. We observed that 
panel discussion was an extremely effective way for motivating participants 
towards research.

\section{Concluding remarks}

We understand that  the workshops have indeed inspired 
some students
to seriously consider algorithmic research as a career option.
In fact, professors of the various host institutes have informed us that
some of their students, who are now pursuing Ph.D. program in computer
science (in India or abroad), were actually motivated by our workshops.
Moreover, 
the benefits from the exposure 
to the broad field of graph and geometric algorithms will certainly get reflected
in the future profession of many of the participants. 
As organizers, we felt that it was worth making the effort, and
we hope that in the interest of the nation and in the interest
of algorithmic studies (in which India was a pioneer), such efforts
will continue in the future.

\newpage

\section*{Appendix}
\appendix
\renewcommand{\thesubsection}{\Alph{subsection}}
\label{tata14102013}
\begin{center}
{\bf Table: Invited Speakers and Lecture Topics}

\bigskip
\footnotesize
\begin{tabular}{| l| l |l| l|}
\hline
Invited speaker& Lecture title& Delivered at\\
\hline
John Augustine  &  (i) Algorithms for Big Data   &  VNIT Nagpur, IIITM\\
IIT Madras&(ii) Introduction to Distributed Algorithms&Kerala, BITS Goa\\
\hline

Amitabha Bagchi& (i) Information Diffusion on Social Networks  & BESU Shibpur, NIT\\
IIT Delhi& (ii) Random Geometric Graphs and Wireless &Patna, SMIT Sikkim \\
&Networks&BHU Banaras\\
\hline

Niranjan  &  (i) Graph Coloring  & University of Kashmir\\
Balachandran&&BESU Shibpur\\
IIT Bombay&&DAIICT Gandhinagar\\
&&IIITDM Jabalpur\\ 
\hline

Arnab Basu  &  (i) Probability and Graphs &  Thapar University\\  
IIM Bangalore& (ii) Big Data: The Future of Computing  & SMIT Sikkim, NIT\\
&& Rourkela\\
\hline

Surender Baswana &  (i) Shortest Paths in Presence of Vertex Failures  & NIT Patna\\
IIT Kanpur & (ii) Algorithms for Graph Spanners: Static, &IISc Bangalore\\ 
&Dynamic and Fault Tolerant&\\
\hline

Amitava& (i) Graph Matching and Applications&PSGTECH Coimbatore\\
Bhattacharya&&IIT Guwahati\\
TIFR Mumbai&&\\
\hline

Partha  Bhowmick&  (i) On Anomalies and Algorithms Related to &NIT Rourkela, NIT\\
IIT Kharagpur&Circles and Spheres in the Digital Space &Warangal, BESU\\
&  (ii) Geometry, Vision, and Graphics: A Number  & Shibpur, NIT Patna\\
& Theoretic Introduction (iii) Anatomies of Lines&IIT Kharagpur, IIT\\
& and Circles in the Light of Number Theory& Guwahati, IIITDM\\ 
&(iv)   Isothetic Covers for Digital Objects: & Jabalpur\\
&Algorithms and Applications&\\ 
\hline

Arijit Bishnu  &  (i) Art Locally, Change Globally: Shape &  Thapar University\\ 
ISI Kolkata& Transformations by Local Interchanges   & NIT Rourkela, NIT\\ 
& (ii) Introduction to Randomized Algorithms  & Warangal, BESU\\
&(iii) Introduction to Network Flows &Shibpur, IIT Kharagpur\\
&(iv) Introduction to Computational Geometry &BHU Banaras, IIITDM\\
&&Jabalpur, IIT Guwahati\\ 
&& SMIT\&SGC Sikkim*\\
\hline
Subhashis Banerjee&  (i) On Large Scale 3D Reconstruction from   & University of Kashmir\\
IIT Delhi & Images and Videos, (ii) Projective Geometry &Thapar University\\
& for Graphics and Computer Vision,  (iii) Singular&  NIT Rourkela, NIT \\
&  Value Decomposition and its Applications to   & Warangal, BESU\\
      &Computer Vision& Shibpur, SGC\\
&&Sikkim, BHU Banaras\\   
&&IIT Guwahati, IIT\\ 
&&Roorkee\\                      
\hline

Sharat Chandran& (i) Visibility Maps for Point Clouds, (ii) Using& CRCE Mumbai, IIT\\
IIT Bombay&Shape Spaces for Structure for Motion,  &Kharagpur, DAIICT\\ 
& (iii) Geometric Data Structures Random &Gandhinagar, IIITM\\ 
& &Kerala\\
\hline

Sunil Chandran & (i) Geometric Representations of Graphs  &  NIT Tiruchirapalli\\
IISc Bangalore&(ii) Rainbow Coloring of Graphs&NIT Rourkela, BITS\\
&& Pilani, BHU Banaras\\
&&BITS Goa\\

\hline

\end{tabular}
\end{center}
* indicates that the speaker had delivered two lectures in this workshop.

\newpage

\begin{center}

\bigskip
\footnotesize
\begin{tabular}{| l| l |l| l|}
\hline
Invited speaker& Lecture topics& Delivered at\\
\hline

Parag Chaudhuri&  (i) Motion Graphs for Character Animation &  Thapar University\\ 
IIT Bombay &&BESU Shibpur, \\ 
&&IIT Guwahati, IIITM\\ 
&&Kerala\\

\hline

Subhasis  Chaudhuri &  (i) Haptic Rendering: How do we touch an & BESU Shibpur, NIT\\
IIT Bombay&object?  &Karnataka, IIT\\ 
&&Roorkee, SMIT Sikkim\\
\hline

Sandip Das  & (i) Introduction to Approximation Algorithms & University of Kashmir\\
ISI Kolkata&  (ii) Convexity of Point Sets  & BESU Shibpur, VNIT\\
&(iii) Geometry Facility Location Problems& Nagpur, NIT Rourkela\\
&(iv) Geometric Approximation Algorithms&NIT Warangal, NIT\\
&&Tiruchirapalli, NIT\\ 
&&Patna, IIT Kharagpur\\
&&IISc Bangalore, BHU\\
&& Banaras, IIT Guwahati\\
&&IIITDM Jabalpur\\
&&IIITM Kerala\\
\hline
Srikrishnan& (i) Introduction to Randomized Algorithms& DAIICT Gandhinagar\\
Divakaran, DAIICT&&\\
Gandhinagar&&\\
\hline 
Ajit  Diwan  &  (i) Extremal Graph Theory, (ii) Graph  &  NIT Tiruchirapalli\\
IIT Bombay&Partitioning, (iii) Topological Minors of Graphs &IISc Bangalore\\
&&PSGTECH Coimbatore\\
&&NIT Karnataka\\
\hline

Sumit Ganguly  & (i) Streaming and Semi-streaming Algorithms for&BHU Banaras\\ 
IIT Kanpur&Processing Massive Graphs, (ii) Introduction to & IIITDM Jabalpur  \\
&Streaming Algorithms&\\

\hline
Deepak Garg  & (i) Introduction to Approximation Algorithms  & Thapar University\\
Thapar University&&\\
\hline
Daya Gaur  &  (i) Approximation Algorithms and Linear & Thapar University\\
IIT Ropar&Programming, (ii) Introduction to Approximation  &IISc Bangalore\\
&Algorithms&DAIICT Gandhinagar\\

\hline

Subir Ghosh &(i) Computational Geometry, (ii) Robot Path & All workshops except\\
TIFR Mumbai&   Planning, (iii) Robot Path Planning: Offline and &at IIT Guwahati. \\
&  On-line Algorithms, (iv) Robot Online Algorithms & (Two lectures were \\
&  for Searching and Exploration in the Plane  & delivered at PSGTECH\\
&  (v) Online Algorithms in Computational Geometry  & Coimbatore.) \\
& (vi) Art Gallery Problems and Approximation &\\
& Algorithms, (vii) Introduction to Network Flows&\\ 
&(viii) Introduction to Approximation Algorithms&\\

\hline

Partha Goswami  &  (i) Introduction to Computational Geometry & University of Kashmir* \\
Calcutta  University&  (ii) Geometric Data Structures  & Thapar University\\
& (iii) Duality Transformation in Geometry  & BESU Shibpur, VNIT \\ 
&& Nagpur, NIT Rourkela\\
&&NIT Warangal, NIT\\
&&Patna, SMIT Sikkim\\
&&NIT Tiruchirapalli\\
&&CRCE Mumbai, IIT\\
&&Kharagpur, BITS\\
&&Pilani, IISc  Bangalore\\
&&BHU Banaras, IIT\\ 
&&Guwahati, NIT\\
&& Karnataka, IIITDM\\ 
&&Jabalpur*, IIT Roorkee\\
\hline

\end{tabular}

\end{center}
* indicates that the speaker had delivered two lectures in this workshop.

\newpage
\begin{center}

\bigskip
\footnotesize
\begin{tabular}{| l| l |l| l|}
\hline
Invited speaker& Lecture topics& Delivered at\\
\hline
Sathish Govindarajan&  (i) Geometric Graphs  & University of Kashmir\\
IISc Bangalore &(ii) Introduction to Combinatorial Geometry& NIT Tiruchirapalli \\
&&NIT Rourkela, NIT\\
&&Patna, BITS Pilani\\
&&IISc Bangalore\\
&&PSGTECH Coimbatore\\
&&NIT Karnataka\\
&&DAIICT Gandhinagar\\
&&IIT Roorkee\\
\hline
Suresh Hegde& (i) Labeled Graphs and Digraphs:Theory and & NIT Karnataka\\
NIT Karnataka&Applications&\\
\hline

R. Inkulu &   (i) Finding Minimum Degree Spanning and    &  VNIT Nagpur, NIT\\
IIT Guwahati& Steiner trees, (ii) Buy-at-Bulk Network Design &  Warangal, NIT Patna\\
& (iii) Graph and Geometric Shortest Paths  &IIT Guwahati\\

\hline

Subrahmanyam  & (i) Introduction to Randomized Algorithms & University of Kashmir\\ 
Kalyanasundaram&&VNIT Nagpur, IIITM\\ 
IIT Hyderabad&&Kerala, IIT Roorkee\\

\hline

Deepak Kapur& (i) Algorithms for Automated Reasoning and & IIT Roorkee\\
UNM, Albuquerque&Symbolic Computations&\\
\hline
K. Murali Krishnan& (i) Graph Representation of Codes and Decoding& IISc Bangalore\\
NIT Calicut& Algorithms&\\
\hline
Subodh Kumar  & (i) Accurate and Efficient Rendering of Detail& VNIT Nagpur, NIT\\
IIT Delhi&using Directional Distance Maps, (ii) Cache &Rourkela, NIT Patna \\
&  Friendly Compressed Representation of Geometry   & IIT Kharagpur, BITS \\
& (iii) Symmetry and Structure Detection for 3D &  Pilani, BITS Goa\\
&Geometry&DAIICT Gandhinagar\\
&&IIT Roorkee\\
\hline

Anil Maheshwari& (i) Geometric Spanners&CRCE Mumbai\\
Carleton University  &&IISc Bangalore\\
Ottawa&&\\
\hline

S.N Maheshwari& (i) Network Flows and Applications&IIT Guwahati\\
IIT Dehi&&\\
\hline

Amitabha Mukerjee &  (i) Robot Motion Planning, (ii) Geometric & University of Kashmir\\
IIT Kanpur &   Modeling for Shape Classes, (iii) Manifold &  Thapar University\\
& Discovery in High-Dimensional Data &NIT Warangal,  SMIT\\ 
&&Sikkim, DAIICT\\ 
&&Gandhinagar, IIITDM\\ 
&&Jabalpur, BITS Goa\\
&&IIT Roorkee\\

\hline

Niloy Mitra& (i) Scalar Field Visualization, (ii) Symmetry& IIT Kharagpur\\
IIT Delhi&and Structure Detection for 3D Geometry&BITS Pilani\\ 
\hline

Anurag Mittal& (i) Graph-based Algorithms in Computer Vision&PSGTECH \\
IIT Madras&&Coimbatore\\
\hline
Sudhir Mudur& (i) The 3D Surface Reconstruction Problem and&BITS Goa\\ 
Concordia University&Some Solutions&\\
Montreal&&\\ 
\hline
Jayanta Mukerjee& (i) In the Quest for Good Digital Distances& IIT Kharagpur\\
IIT Kharagpur&&\\

\hline
Krishnendu Mukerjee& (i) Distributed Leader Election  & NIT Patna\\
ISI Kolkata&&\\
\hline

Subhas Nandy& (i) Introduction to Randomized Algorithms  & Thapar University\\
ISI Kolkata&(ii) Voronoi Diagram& NIT Rourkela, CRCE\\
&&Mumbai,  BITS Pilani\\
&& IIT Kharagpur, IISc\\
&&Bangalore, BHU\\
&&Banaras\\
\hline

\end{tabular}

\end{center}

\newpage
\begin{center}

\bigskip
\footnotesize
\begin{tabular}{| l| l |l| l|}
\hline
Invited speaker& Lecture topics& Delivered at\\
\hline

P. Narayanan& (i) Graphics Processor Units: For Graphics and&IISc Bangalore\\
IIIT Hyderabad&Beyond&\\
\hline

Vijay Natarajan&(i) Scalar Field Visualization: Level Set Topology & IIT Kharagpur, BITS\\ 
IISc Bangalore& (ii) Symmetry in Scalar Fields &Pilani, IISc Bangalore\\
 &&PSGTECH Coimbatore\\
&&NIT Karnataka, BITS\\ 
&&Goa, NIT Nagpur\\

\hline
N. S. Narayanaswamy&  (i) Special Classes of Intersection Graphs  &  NIT Tiruchirapalli\\
IIT Madras& (ii) Perfect Graphs, (iii) Graph Classes with & NIT Warangal, IISc\\
&Interesting Structural Properties, (iv) Tree Path& Bangalore, IIT\\ 
&Assignments: An extension of the Consecutive &Guwahati, NIT\\
&Ones Property, (v) Algorithms for Perfect Graphs&Karnataka, IIITDM\\ 
&&Jabalpur, IIT Roorkee\\

\hline
Sudebkumar Pal  & (i) Link Paths and Reflection Visibility   &  NIT Rourkela, SMIT\\
IIT Kharagpur& Problems, (ii) Geometric Data Structures& Sikkim, NIT Patna\\ 
&&NIT Tiruchirapalli\\
&&CRCE Mumbai, BITS\\
&&Pilani, IISc  Bangalore\\
&&BHU Banaras, IIT\\ 
&&Guwahati, NIT\\ 
&&Karnataka, DAIICT\\ 
&&Gandhinagar, IIITDM\\ 
&&Jabalpur, IIITM\\ 
&&Kerala, IIT Roorkee\\

\hline

Sachin Patkar& (i) Network Flows and Applications& BITS Goa\\
IIT Bombay&&\\
\hline
Shariefuddin Pirzada&  (i) Graph Spectra and Applications  & University of Kashmir\\
UK Srinagar&&\\

\hline

Venkatesh  Raman& (i) Fixed Parameter Algorithms  &  NIT Tiruchirapalli\\
IMSc Chennai&&IISc Bangalore\\
&&PSGTECH Coimbatore\\
\hline

Abhiram Ranade  &  (i) Graph Partitioning, (ii) Geometric Packing&  VNIT Nagpur,  NIT\\
IIT Bombay&(iii) Algorithms for Precedence Constrained & Warangal, BITS Goa \\ 
&Scheduling, (iv) Mumbai Navigator, (v) Some &BESU Shibpur, SMIT \\ 
& Formulations of the Genome Assembly Problem &Sikkim, BITS Pilani \\
& & PSGTECH Coimbatore\\
&& DAIICT Gandhinagar*\\ 
&&BHU Banaras, IIITM\\ 
&&Kerala, IIT Roorkee\\

\hline
M. Panduranga Rao& (i) Introduction to Quantum Algorithms&BITS Goa\\
IIT Hyderabad&&\\

\hline

Tathagata Ray  &  (i) Mesh Generation  &  NIT Warangal, IIITM\\ 
BITS Hyderabad&&Kerala\\
\hline
Sasanka Roy &  (i) Introduction to Computational Geometry  &  VNIT Nagpur, IIT\\
CMI Chennai&(ii) Voronoi Diagram& Guwahati, PSGTECH\\ 
&&Coimbatore, BITS Goa\\

\hline

Yogish Sabharwal& (i) Randomized Techniques in Geometry& IIT Kharagpur\\ 
IBM Delhi&&\\

\hline
Sudeep Sarkar& (i) Graphs and Object Recognition&IISc Bangalore\\
USF, Tampa&&\\

\hline
\end{tabular}

\end{center}
* indicates that the speaker had delivered two lectures in this workshop.
\newpage
\begin{center}

\bigskip
\footnotesize
\begin{tabular}{| l| l |l| l|}
\hline
Invited speaker& Lecture topics& Delivered at\\
\hline

Swami   & (i) Planarity Testing of Graphs, (ii) Voronoi & Thapar University\\
Sarvattomananda&Diagrams, (iii) Geometric Data Structures &BESU Shibpur,  NIT\\     RKMVU, Belur & (iv) Helly's Theorem and Centre Point &  Rourkela, NIT Patna\\ 
&(v) Computing Center Point Using the Prune&SGC Sikkim, CRCE\\
& and Search Technique, (vi) Prune and search   &Mumbai, BITS Pilani\\
&Technique in Geometry&IIT Kharagpur, IISc\\
&& Bangalore, BHU\\ 
&&Banaras, PSGTECH\\ 
&&Coimbatore, IIITM\\ 
&&Kerala\\

\hline
Saket Saurabh  &  (i) Fixed Parameter Algorithms  &  NIT  Rourkela, NIT\\
IMSc Chennai&(ii) Preprocessing with Guarantee  & Patna, NIT Karnataka\\

\hline
Sandeep Sen& (i) A Short Guided Tour of Randomized & BITS Pilani\\
IIT Delhi&Algorithms&IISc Bangalore\\
\hline
Naveen Sivadasan&  (i) Introduction to On-line Algorithms   & Thapar University\\
IIT Hyderabad&&VNIT Nagpur, NIT\\ 
&& Warangal, NIT\\ 
&&Karnataka\\

\hline

K. V.  Subrahmanyam &(i) Expander Graphs and their Applications  & NIT Tiruchirapalli\\
CMI Chennai&(ii) Network Flows and Applications   & NIT Patna\\

\hline
C. R. Subramanian   &  (i) Introduction to Randomized Algorithms  & NIT Tiruchirapalli\\
IMSc Chennai& (ii) Randomized Algorithms for Counting & BHU Banaras, IISc\\
&Problems&Bangalore, PSGTECH\\
&&Coimbatore, NIT\\ 
&&Karnataka, BITS Goa\\
\hline
Ambat Vijayakumar& (i) Graph Dynamics&IIITM Kerala\\
CUSAT, Cochin&&\\
\hline
Nisheeth Vishnoi& (i) Hardness of Approximation& NIT Karnataka\\
Microsoft Bangalore&&\\
\hline

\end{tabular}

\end{center}

\end{document}